\documentclass{article}
\usepackage{amsmath,amsthm,amssymb,amsfonts,mathrsfs, fancyhdr, color, comment, graphicx, environ, caption, subfigure, float}
\usepackage[a4paper, scale=0.7]{geometry}
\usepackage[backend=biber, style=authoryear, citestyle=authoryear]{biblatex}
\usepackage[hidelinks, colorlinks=true, linkcolor=blue, citecolor=blue, urlcolor=blue]{hyperref}
\usepackage{authblk}
\usepackage{hyperref}
\hypersetup{colorlinks=true}
\title{\textbf{State-dependent Filtering of the Ring Model}}
\author[1]{Jing Yan}
\author[1]{Yunxuan Feng}
\author[2,3]{Wei Dai\thanks{Corresponding author. Email: weidai@fudan.edu.cn}}
\author[1]{Yaoyu Zhang\thanks{Corresponding author. Email: zhyy.sjtu@sjtu.edu.cn}}
\affil[1]{School of Mathematical Sciences, Institute of Natural Sciences and MOE-LSC, Shanghai Jiao Tong University, Shanghai, China}
\affil[2]{Fudan University, Research Institute of Intelligent Complex Systems, Shanghai, China}
\affil[3]{Shanghai Artificial Intelligence Laboratory, Shanghai, China}

\begin{document}

\maketitle
\section*{Abstract}
Robustness is a measure of functional reliability of a system against perturbations. 
To achieve a good and robust performance, a system must filter out external perturbations by its internal priors. 
These priors are usually distilled in the structure and the states of the system. 
Biophysical neural network are known to be robust but the exact mechanisms are still elusive.
In this paper, we probe how orientation-selective neurons organized on a 1-D ring network respond to perturbations in the hope of gaining some insights on the robustness of visual system in brain.
We analyze the steady-state of the rate-based network and prove that the activation state of neurons, rather than their firing rates, determines how the model respond to perturbations. 
We then identify specific perturbation patterns that induce the largest responses for different configurations of activation states, and find them to be sinusoidal or sinusoidal-like while other patterns are largely attenuated.
Similar results are observed in a spiking ring model.
Finally, we remap the perturbations in orientation back into the 2-D image space using Gabor functions. 
The resulted optimal perturbation patterns mirror adversarial attacks in deep learning that exploit the priors of the system.
Our results suggest that based on different state configurations, these priors could underlie some of the illusionary experiences as the cost of visual robustness.


\section*{Keywords}
ring model, robustness, state-dependent filtering, visual processing

\section{Introduction}
Robustness, the ability of a system to maintain its functionality against perturbations, is a critical property for many complex systems, including neural networks (\cite{RN219,RN220}). The visual system, in particular, exhibits remarkable robustness, accurately perceiving and recognizing objects despite variations in lighting, viewpoint, and other distortions in stimuli (\cite{RN218}). Understanding the mechanisms underlying this robustness is a key goal in computational neuroscience and has important implications for developing robust artificial visual systems.

One promising approach to studying robustness in neural networks is through the analysis of simplified models that capture essential features of biological networks while remaining analytically tractable. The ring model, a recurrently connected network of orientation-selective neurons with Gaussian-shaped connectivity on a 1-D ring, has been widely used to study various aspects of visual processing, including orientation selectivity (\cite{RN221}), contrast invariance (\cite{RN222}), surround suppression (\cite{RN223,RN224,RN241}), and even for binocular rivalry and fusion(\cite{RN242,RN243,RN244}).

Despite its simplicity, such models can incorporate effective single-neuron nonlinearity and connectivity from experimental data that determine the possible configurations of neuronal states that approximate a real visual system.
The space spanned by the the states could be considered as the priors of the model with respect to the external stimulus.
Thus, the properties of these priors are what underlie the robustness of a model against various perturbations.
Specifically, the model's response to perturbations depends on the configuration of neuronal activation states that arises from the nonlinear activation function and the interactions between neurons.

In this paper, we investigate such state-dependent response of the ring model to perturbations and its potential implications for the robustness of the visual system. 
We first analyze a simplified steady-state ring model and derive analytical expressions that relate the responses of several models with specific connectivity to its activation state and the external perturbation.
We then identify the perturbations that induce the largest responses for different activation states and explore their properties. 
In addition, we examine such state-dependent response in a more biologically plausible spiking network model and compare the results to the steady-state rate model.
Finally, we remap the response of the models in the orientation domain to a 2-D image domains through Gabor functions.

Our findings provide insights on the mechanism of robust visual processing in brain through an ring model with effective nonlinearity and connectivity. By elucidating the model's state-dependent responses to perturbations, we instantiate the priors that filter external perturbations and paves the way for developing more robust and biologically inspired artificial vision systems.

\section{Results}
\subsection*{The perturbed system: determined by lateral connections and activation pattern}
The ring model is a simplified model of the orientation columns found in the primary visual cortex (V1) where orientation preferences are indicated by the polar angles on the ring. It consists of two populations of neurons, representing excitatory (E) and inhibitory (I) neurons uniformly placed on a ring and laterally connected through a Gaussian-shaped connectivity kernel. We first study the steady-state rate ring model (See Section~\ref{sec:materials} \nameref{sec:materials}.), which satisfies the following equation (Eq. \ref{eq:steady_rate}).
\begin{equation} \label{eq:steady_rate}
\begin{aligned}
r_{E} &= g(I_{E} + k_{EE}\ast r_{E} - k_{EI}\ast r_{I})\\
r_{I} &= g(I_{I} + k_{IE}\ast r_{E})
\end{aligned}
\end{equation}
Here, $r_{X}$ ($X\in\{E, I\}$, same for the $Y$ latter) denotes the firing rate vector of the neuron population. $I_{X}$ represents the external input, and $k_{XY}$ is the connectivity kernel from population $Y$ to $X$, implementing the Gaussian profile (see Section~\ref{sec:materials} \nameref{sec:materials}). The activation function $g$ is chosen as the rectified linear unit (ReLU), and $\ast$ denotes the circulant convolution operation, enforcing the ring topology. 

To investigate the model's response to perturbations, we derived the linearized system around the steady-state solution (for any inputs) by considering small perturbations $\delta I_{X}$.

\begin{equation} \label{eq:perturbed_rate}
\begin{aligned}
\delta r_{E} &= g^{\prime}_{E}\odot(\delta I_{E} + k_{EE} * \delta r_{E} - k_{EI} * \delta r_{I})\\
\delta r_{I} &= g^{\prime}_{I}\odot(\delta I_{I} + k_{IE} * \delta r_{E})\\
\end{aligned}
\end{equation}

Here, $g_{X}^{\prime}$ represents the derivation of the activation function for population $X$, and $\odot$ denotes element-wise multiplication. The perturbation should be small enough so that neurons active state are not changed. Crucially, since the ReLU activation is piecewise linear, $g_{X}^{\prime}$ depends only on which neurons are active (i.e., have non-zero firing rates), and not on the specific firing rate values.
Note that the perturbed rate equation have shown that the model's local behavior does not directly depend on the inputs $I$, but on the $g^{\prime}$. This observation reveals that the ring model's response to perturbations depend only on neurons active state. The number of possible response patterns scaling as $2^{n}$, where $n$ is the number of neurons, bringing a reasonably rich diversity of behaviors against perturbations. 

In the following sections, we analyze this state-dependent response in detail, identifying the perturbations that induce the largest responses for different activation states and exploring their properties in domains of both neural response and the effectively reconstructed input.

\subsubsection*{Effects of lateral connections on model response with fully active neurons}
We first consider the case that all neurons are active.
The perturbed system can be solved in the frequency space, which implies that sinusoids are eigenvectors in this case. A sinusoidal signal retains its shape but only changes in intensity as it passes through the system. The change in intensity, we call it gain from now on, is related to the frequency of the signal itself and is determined by the lateral connections. The specific relation is given by
\begin{equation} \label{eq:rE solution in freq}
    \hat{\delta r}_{E} = (1- \hat{k}_{EE} + \hat{k}_{EI}\hat{k}_{IE})^{-1}(\hat{\delta I}_{E} - \hat{k}_{EI}\hat{\delta I}_{I}).
\end{equation}
where $\hat{v}$ is DFT (Discrete Fourier transform) of a vector $v$. Details are shown in Section~\ref{sec:materials} \nameref{sec:materials}. We denote
\begin{equation} \label{eq:h inv in freq}
    \hat{h}_{-1} := \hat{h}_{0}^{-1} :=
    (1- \hat{k}_{EE} + \hat{k}_{EI}\hat{k}_{IE})^{-1},
\end{equation}
which is the gain across different frequencies.

Note that lateral connections can be divided into two parts:
the excitatory part $\hat{k}_{EE}$ (or $k_{EE}$ in the orientation space) enhancing signals through the lateral connections among excitatory neurons, and the
recurrently inhibitory part $\hat{k}_{EI}\hat{k}_{IE}$ (or $k_{EI}\ast k_{IE}$ in the orientation space) which reduces signals through the combined effect of connections from excitatory neurons to inhibitory neurons and vise versa.
Now we consider possible shape of $\hat{h}_{-1}$ in the assumption that the kernels of lateral connections are Gaussian.

Gaussian kernels remain Gaussian in the frequency space. Thus, the excitatory and recurrently inhibitory parts enhance and reduce low-frequency components in perturbations, respectively, and this enhancement or reduction  weakens gradually as the frequency increases. Sufficiently high-frequency components pass through the system nearly unchanged.   

\begin{figure}[h!]
    \centering
    \includegraphics[width=1\textwidth]{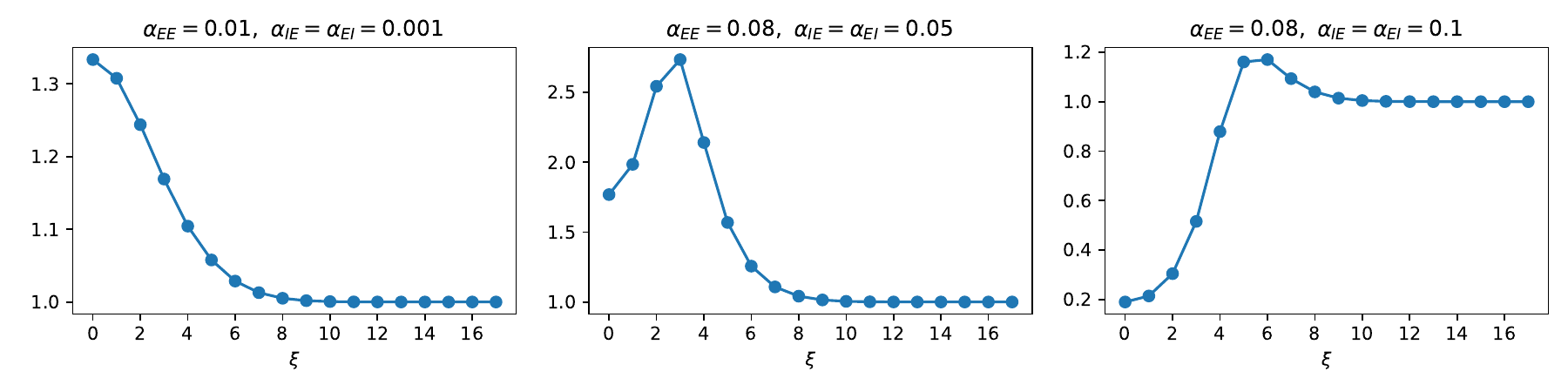}
    \caption{Possible shapes of gain curves are presented, with the horizontal axis representing frequency and the vertical axis representing gain. The connectivity spread between neurons are set to be the same, with only the strength changing. The strength is denoted in the titles as the amplitude of the Gaussian kernel.}
    \label{fig:gain curves of different cncts}
\end{figure}

The range of influenced frequencies is inversely proportional to the connections scope. Since $k_{IE}$ and $k_{EE}$ are usually of same width (same variance in Gaussian functions), the recurrently inhibitory part ($k_{EI}\ast k_{IE}$) has wider scope than the excitatory part ($k_{EE}$). Consequently, the frequencies reduced by the inhibitory part are often less then those enhanced by the excitatory part. The possible shapes of $h_{-1}$ are shown in Figure \ref{fig:gain curves of different cncts}.

\subsubsection*{Effect of activation patterns on model response}
We discuss how the system's behavior changes with differnt activation patterns. 
Our focus is on perturbations that can lead to strong responses. Note that the system remains linear regardless of the activation pattern. So we can find singular vectors of the system (we can always do that by writing the system in a matrix-vector form). Using a set of lateral connections (with Gaussian shapes), we calculate the singular vectors with different activation patterns. We find that these singular vectors are sinusoids or sinusoidal-like signals. By sinusoidal-like, we mean a signal has frequencies concentrated around a particular frequency. Typical examples are shown in Figure \ref{fig:singular vectors, all active}, \ref{fig:singular vectors, first half active}, \ref{fig:singular vectors, keep even, range all}.

\begin{figure}[h!]
    \centering
    \includegraphics[width=1\textwidth]{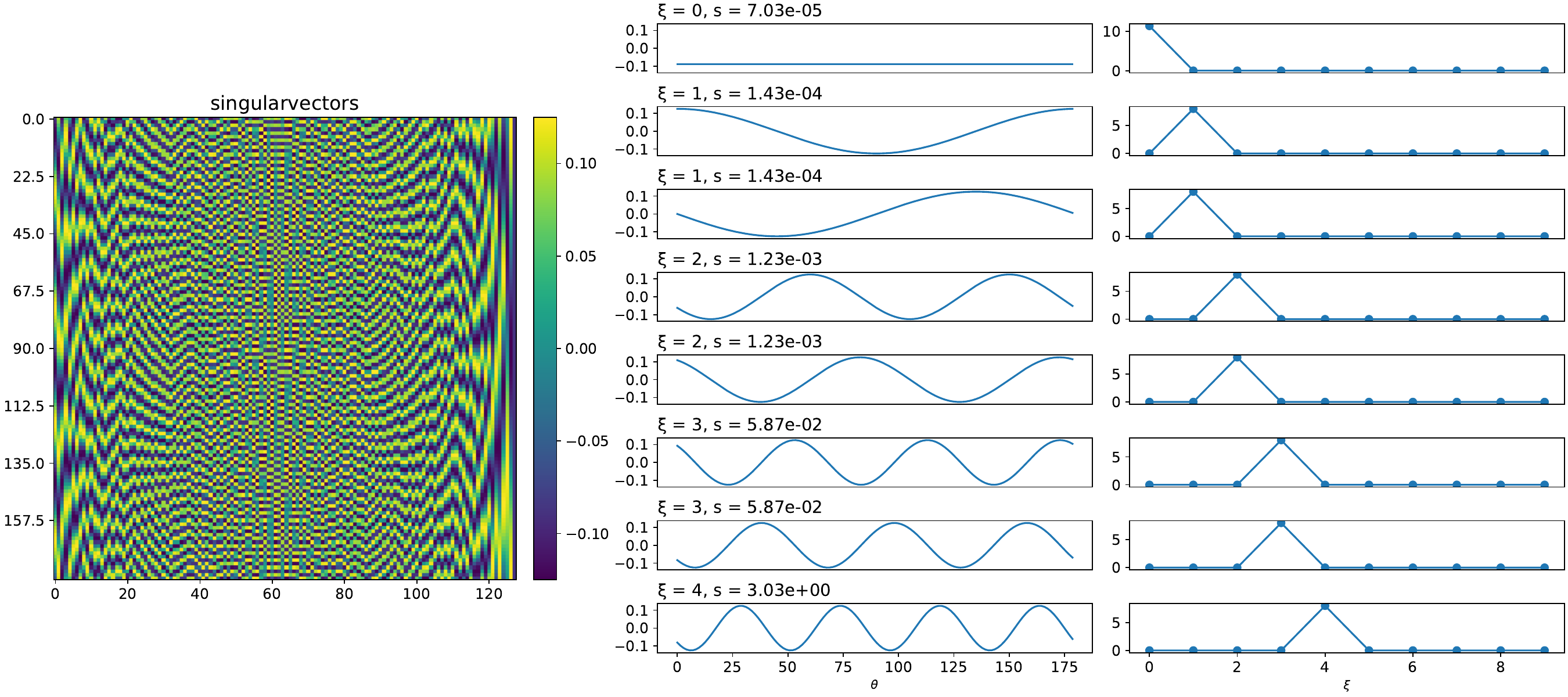}
    \caption{Singular vectors of the perturbed system with all neurons active. The left most is the matrix of the singular vectors. Each column is a singular vector. They are arranged in the order of singularvalues from large to small. The singular vectors are sinusoids and we plot them in original space and frequency space respectively in the middle and right columns, in the order of frequency from small to large.}
    \label{fig:singular vectors, all active}
\end{figure}

\begin{figure}[h!]
    \centering
    \includegraphics[width=1\textwidth]{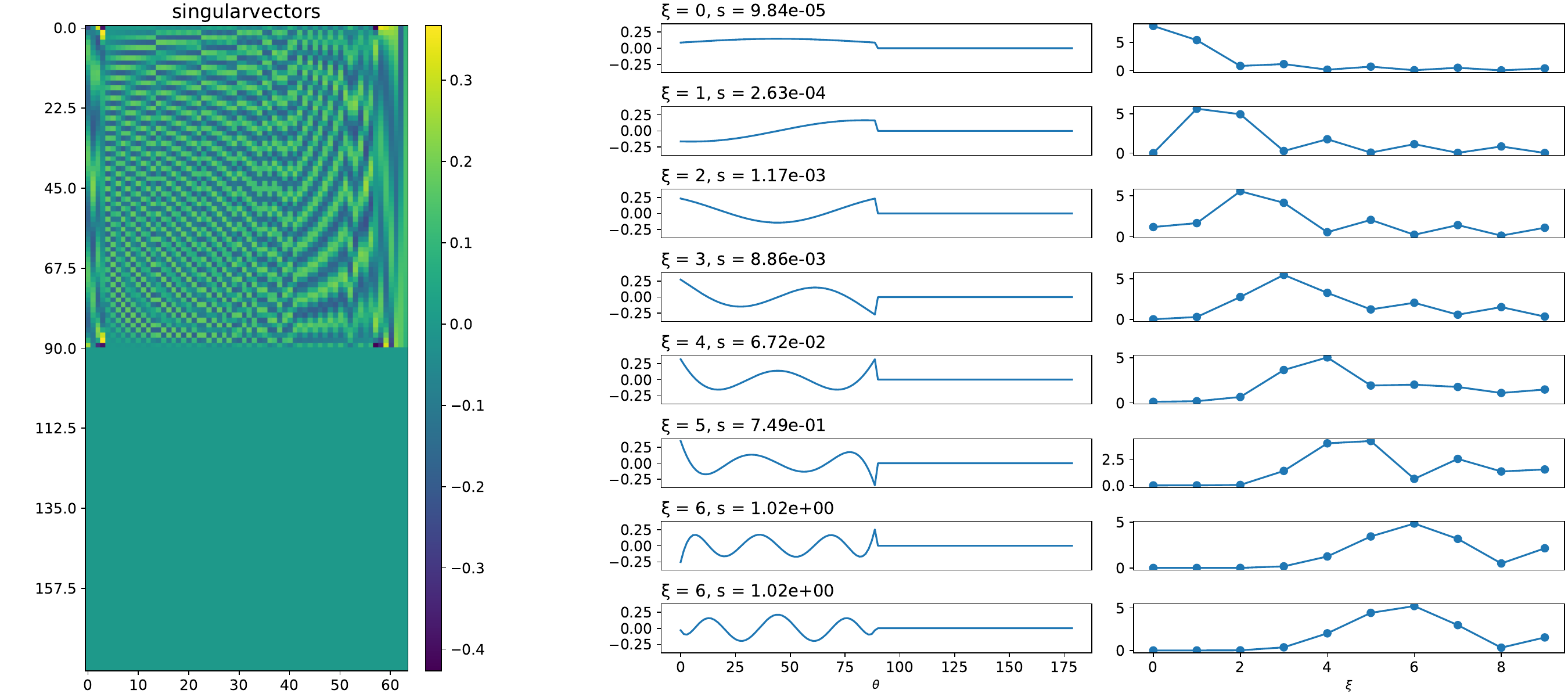}
    \caption{Singular vectors of the perturbed system with half successive neurons active. }
    \label{fig:singular vectors, first half active}
\end{figure}

\begin{figure}[h!]
    \centering
    \includegraphics[width=1\textwidth]{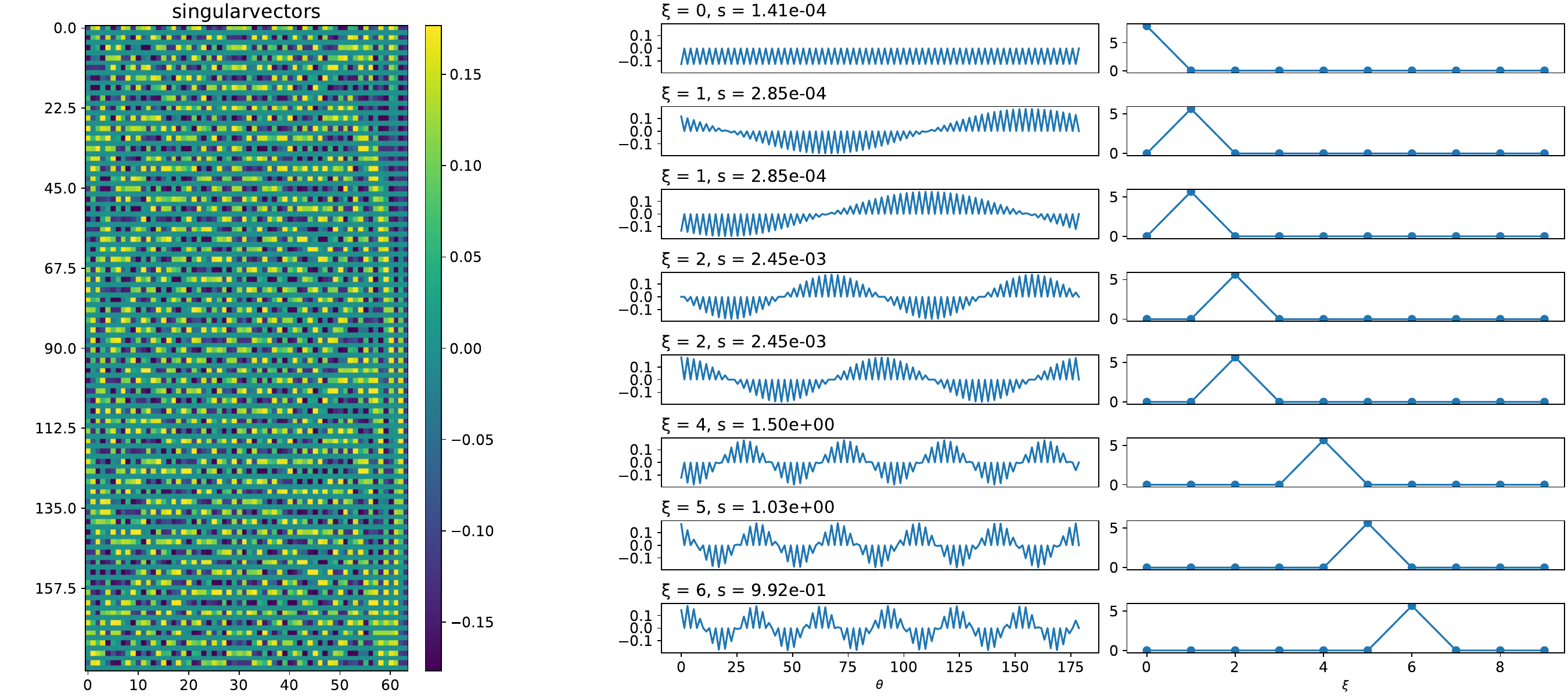}
    \caption{Singular vectors of the perturbed system with neurons being alternatively active.}
    \label{fig:singular vectors, keep even, range all}
\end{figure}

Since the singular vectors closely resemble masked sinusoids, we can plot the gain against different frequencies to compare the effects brought by varying activation patterns.  

\begin{figure}[h!]
    \centering
    \includegraphics[width=1\textwidth]{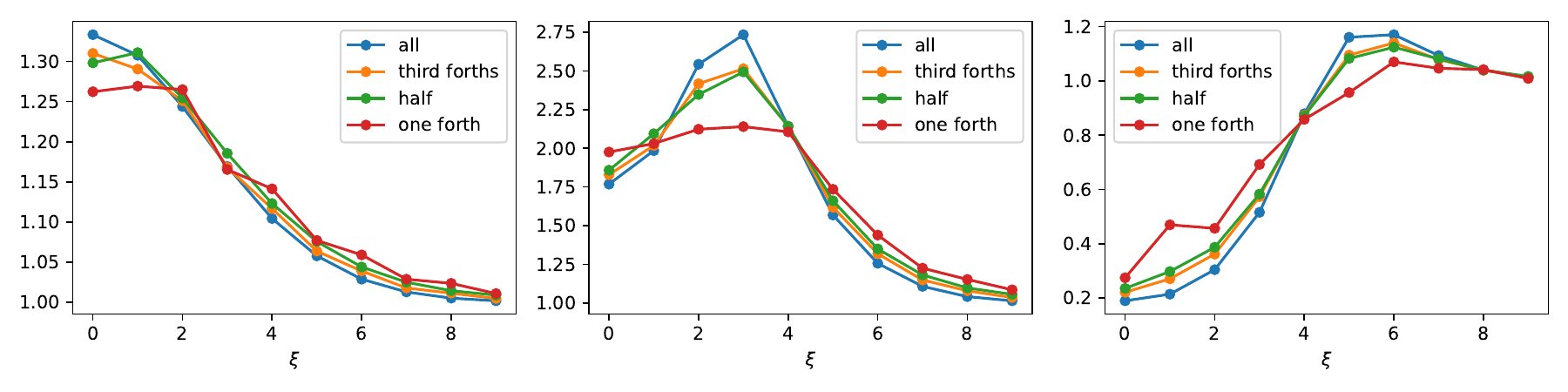}
    \caption{Gain curves of different activation patterns. Three different sets of lateral connections are chosen.}
    \label{fig:gain curves of different activation patterns}
\end{figure}

Figure \ref{fig:gain curves of different activation patterns} illustrates the different gain curves for various activation patterns and lateral connections. We observe that as the activation area shrinks, The gain curve across different frequencies becomes increasingly smoother due to an averaging effect.

\subsection*{Results for spiking ring models}

\begin{figure}[h!]
    \begin{minipage}{0.45\textwidth}
    \centering
    \includegraphics[width=\textwidth]{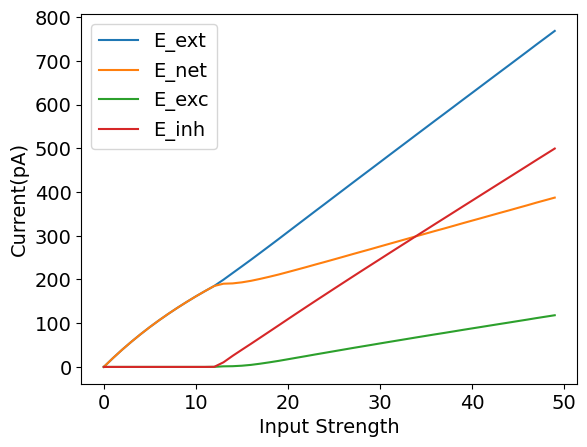}
    \end{minipage}
    \hfill
    \begin{minipage}{0.45\textwidth}
    \centering
    \includegraphics[width=\textwidth]{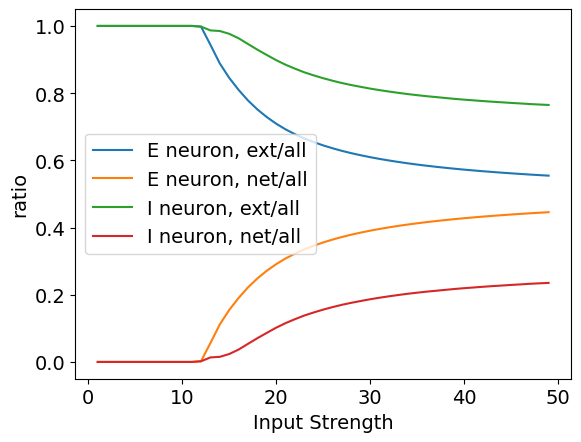}
    \end{minipage}
    \caption{Validation of parameter rationality. The figure on the left illustrates the variation of the network current as the input strength varies. The right figure demonstrates the internal network current versus the external input current as a percentage of the total network current.}
\end{figure}

To validate the generality of our theoretical findings, we conducted experiments using a more biologically realistic conductance-based spiking neuron model. 
While the steady-state rate model and the spiking neuron model differ significantly in their implementations, important terms such as 'firing rates' and 'lateral connections' are preserved across both models (See Section~\ref{sec:materials} \nameref{sec:materials}).

Based on previous work with this model, we considered two parameter regimes: the mean-driven regime and the fluctuation-driven regime (\cite{cai2004effective}). The mean-driven regime is more compatible with our theoretical analysis while still capturing biological characteristics, whereas the fluctuation-driven regime is more aligned with actual biological neuronal responses. We obtained results in the mean-driven regime that fully align with the steady-state rate model, and qualitatively consistent results in the fluctuation-driven regime. In the fluctuation-driven regime, we carefully investigated the differences that led to quantitative deviations as shown in Section \ref{sec:fluctuation driven}.

First, we verified how well the model matches biological phenomena. We demonstrated that the model can reproduce biological observations, such as lateral inhibition and winner-take-all behavior, which are essential for tasks like orientation selectivity.

\begin{figure}[h!]
    \centering
    \includegraphics[width=0.8\textwidth]{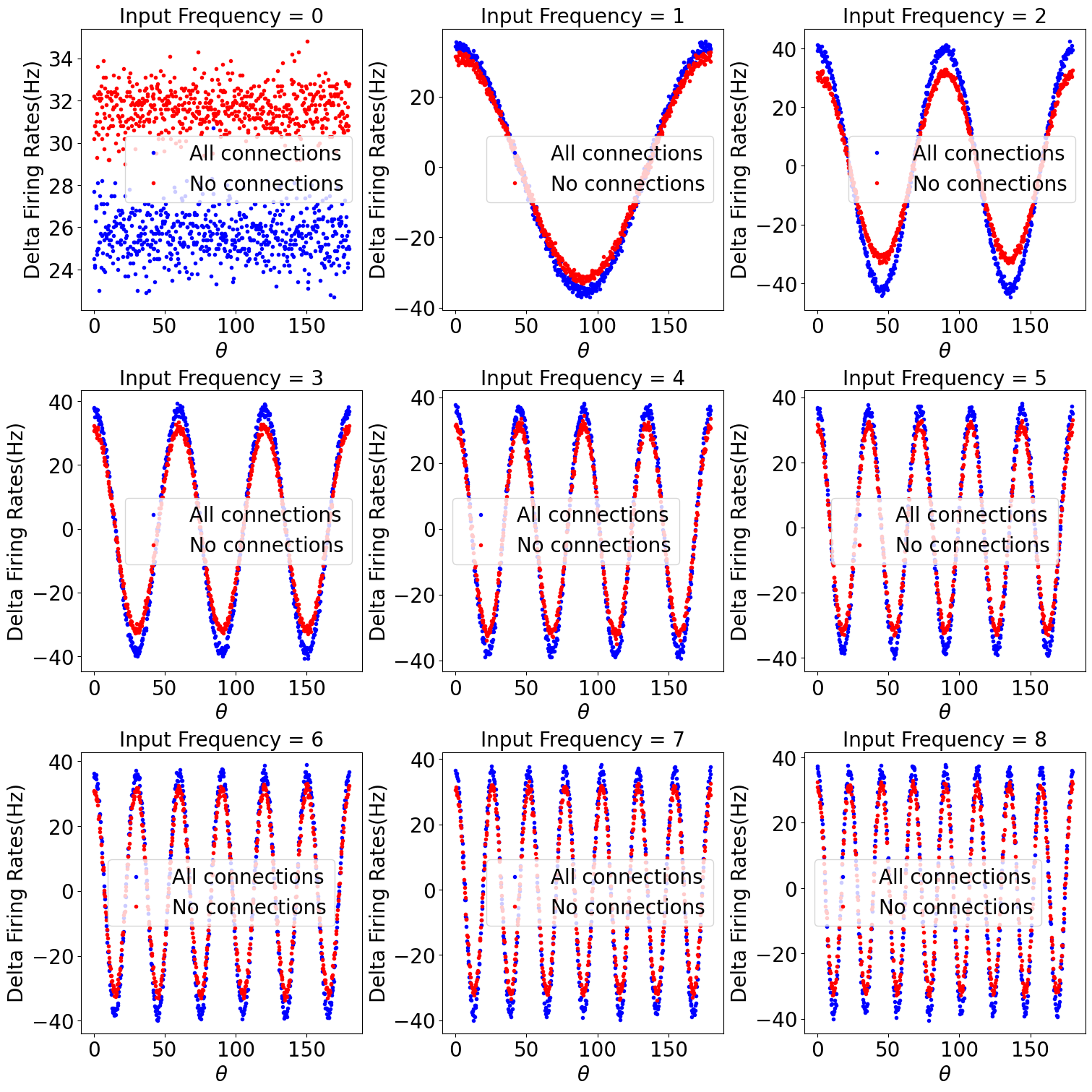}
    \caption{The effect of perturbation on firing rates. This figure shows the change in firing rate brought about by different frequency inputs in the case of full connectivity.}
    \label{fig:changes_in_rates}
\end{figure}

\begin{figure}[h!]
    \centering
    \includegraphics[width=0.8\textwidth]{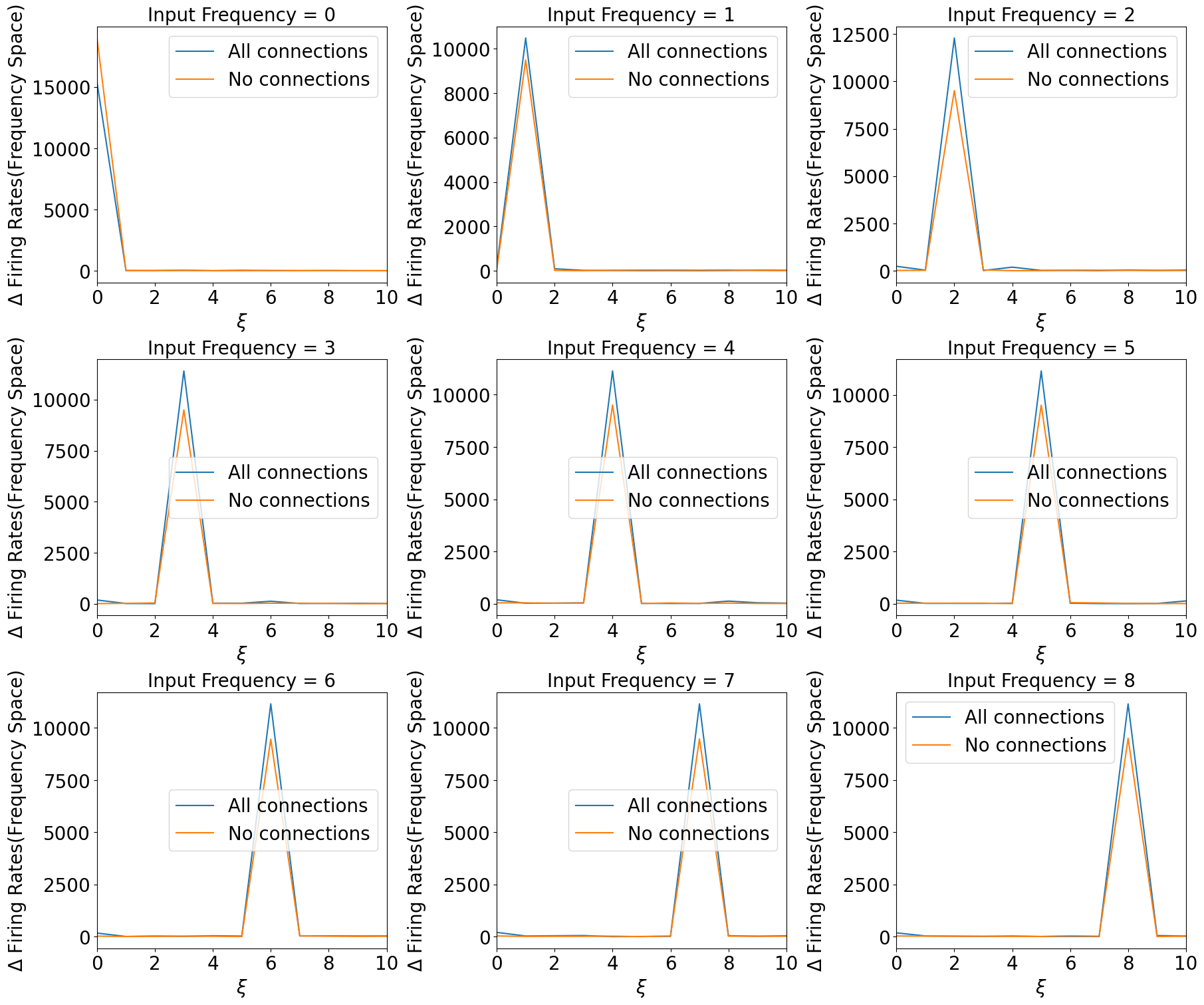}
    \caption{The effect of perturbation on firing rates in frequency space. This figure shows the change in frequency space brought about by different frequency inputs in the case of full connectivity.}
    \label{fig:changes_in_frequency}
\end{figure}

Next, we present experimental results in the mean-driven regime. Consistent with the steady-state rate model, the spiking neuron model in the mean-driven regime also demonstrated a preference for different frequencies. This quantitative consistency could be achieved by considering a dimensionful mapping from the dimensionless steady-state rate model to the dimensionful spiking neuron model.

\begin{figure}[h!]

    \begin{minipage}{0.3\textwidth}
    \centering
    \includegraphics[width=\textwidth]{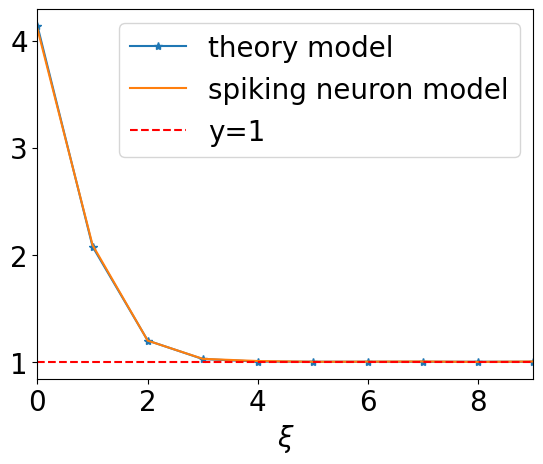}
    \end{minipage}
    \hfill
    \begin{minipage}{0.3\textwidth}
    \centering
    \includegraphics[width=\textwidth]{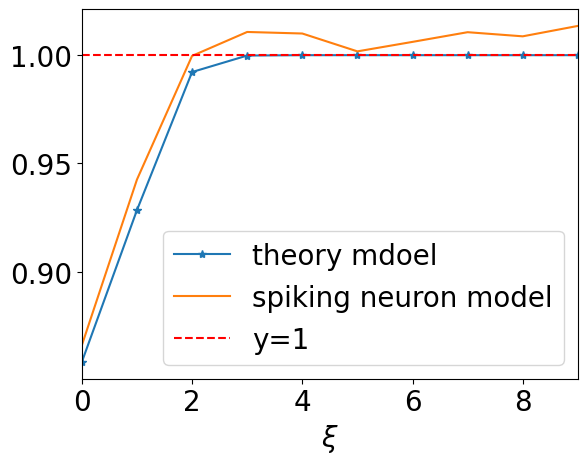}
    \end{minipage}
    \hfill
    \begin{minipage}{0.3\textwidth}
    \centering
    \includegraphics[width=\textwidth]{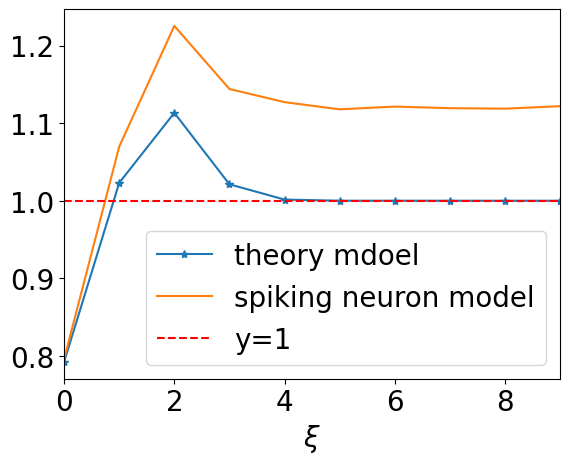}
    \end{minipage}
    \caption{Gain curves of spiking neuron model(mean-driven regime). The figure on the left shows the results with only recurrent excitation connections. The middle figure shows the results with only recurrent inhibition. The right figure demonstrates the results with full connectivity.}
    \label{fig:mean-driven}
\end{figure}

As shown in the Figure \ref{fig:mean-driven}, when no connections were present, the ring model responded uniformly to all frequencies. With only recurrent excitatory  connections, the ring model enhanced the low-frequency response. In contrast, when only recurrent inhibitory connections were present, the ring model suppressed the low-frequency response. And when all connections are present, the model demonstrates a choice of specific frequency.

\subsection*{Remapping perturbation patterns in the image space}
We have discussed how a ring model's filtering property depend on its lateral connections and activation pattern. In particular, we have shown that sinusoidal or sinusoidal-like signals are singular vectors of the perturbed system. Now, we will discuss the image patterns which can correspond to these singular vectors.
To achieve this, we construct a Gabor filter $\mathcal{F}_{\mathcal{G}}$, which is a linear operator mapping from image space to signal space. (See Section~\ref{sec:materials} \nameref{sec:materials}.) 

We analyze the Gabor filter independently to understand its filtering properties. Specifically, we perform singular value decomposition (SVD). The results shows that sinusoids are exactly the (left) singular vectors of the Gabor filters. Consequently, the right singualr vectors are exactly the corresponding patterns of sinusoids in the image space. These patterns are displayed in Figure \ref{fig:patterns in the image space, all active}, and consist of stripes radiating from the center and distributed on a ring. The radius of the ring is proportional to the frequency, as is the number of stripes.

\begin{figure}[h!]
    \centering
    \includegraphics[width=1\textwidth]{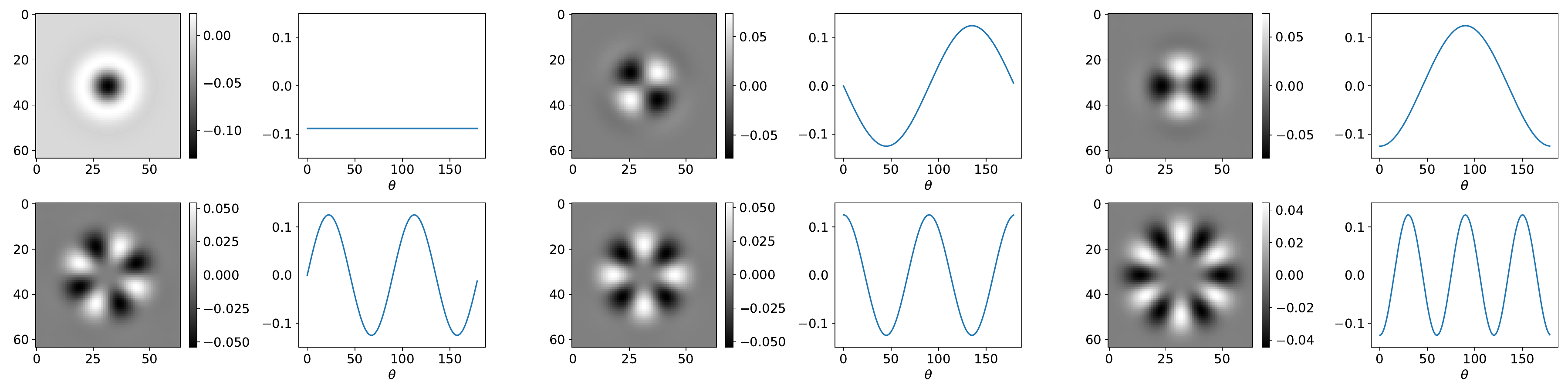}
    \caption{$U$ and $V$ of the gabor filters $\mathcal{F}_{\mathcal{G}}$. Columns of $V$ are reshaped to be images ploted from left to right first and then top to bottom. The corresponding columns of $U$ are plotted as curves on the right side of the images.}
    \label{fig:patterns in the image space, all active}
\end{figure}

When only part of neurons are active, the signals in positions with inactive neurons are filtered. We then take the submatrix from the Gabor filter, composed of rows corresponding to active neurons, and denote it as $\tilde{\mathcal{F}}_{\mathcal{G}}$. The remaining rows also form a matrix, denoted as $\tilde{\mathcal{F}}_{\mathcal{G}}^{c}$
In this scenario, we follow the concept behind SVD and use an optimization algorithm to find the patterns in the image space that are orthogonal to each other and satisfy the constraint $|\tilde{\mathcal{F}}^{c}_{\mathcal{G}}p|_{\infty}\leq| \tilde{\mathcal{F}}_{\mathcal{G}}p|_{\infty}$ (where ($p$) represents a pattern). The patterns should retain the most "energy" (having outputs with the largest norm) while meeting these constraints.

\begin{figure}[h!]
    \centering
    \includegraphics[width=1\textwidth]{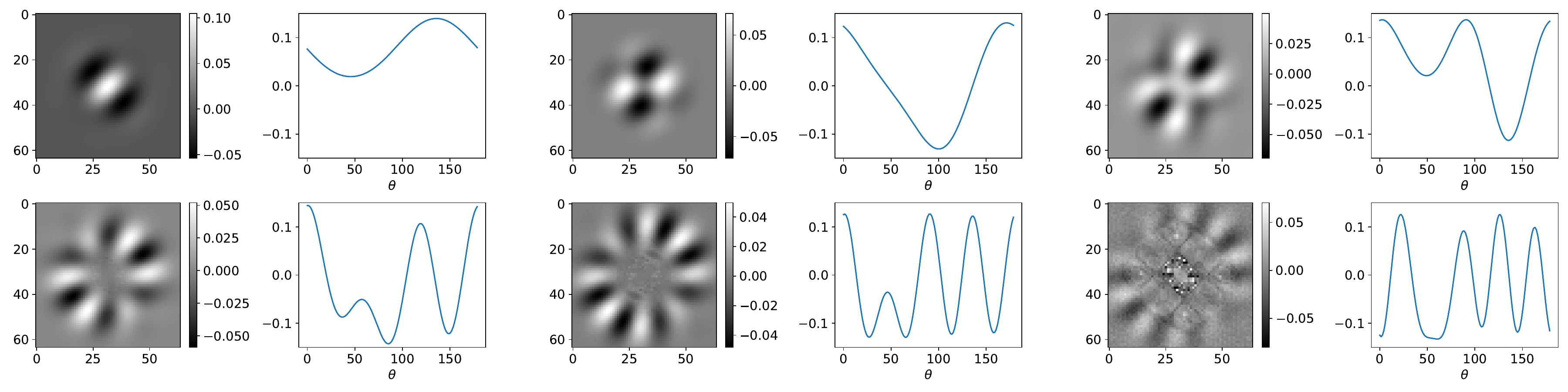}
    \caption{Patterns that orthogonal to each other and the corresponding signals.}
    \label{fig:patterns in the image space, half active}
\end{figure}

As shown in Figure \ref{fig:patterns in the image space, half active}, the patterns exhibit a certain directionality based on the previous patterns.

\section{Discussion} \label{sec:discussion}
\subsection*{Advantages and limitations}
This study provides insights into the robustness mechanisms of the visual system using the ring model, which is a simplification of the network of orientation-selective neurons in the primary visual cortex (V1). One of the key findings is that the system's response to perturbations is highly dependent on the state of neuronal activation and the connectivity between neurons as its internal priors. By employing both a steady-state rate model and a spiking network model, we bridge theoretical understanding and biological plausibility.

The discovery that different activation patterns can alter the filtering properties of the system is particularly important. It suggests that the visual system can adaptively respond to perturbations based on its current state which is thought to involve the context of the external stimuli and the prior of the system. This phenomenon echoes the adaptive nature of biological organisms in real-world situations (\cite{RN218}).
Our analysis reveals that gain variations across different frequencies are dictated by connection strength. This supports the idea that biological neural networks are fine-tuned to respond optimally to environmental inputs while filtering out noise, a concept that could inspire more adaptive artificial neural networks (\cite{RN222}). However, in this work we've only demonstrate such adaptive capacity by analyzing predetermined activation patterns, the mechanisms for a closed-loop adapting process require further investigations.

Additionally, by mapping the transformation from the image space to the signal space, we show that the radius of optimal response pattern corresponds to signal frequency, providing a way to understand the filtering properties. This could form the basis for new approaches in image processing where robustness to distortions is a key requirement (\cite{RN219}).

While our findings advance the understanding of robustness in neural systems, the study has several limitations.
Despite the ring model's analytical tractability, its simplicity may overlook important aspects of V1's functionality. Complex interactions and higher-order visual processing capabilities are not captured, which could limit the generalization of our findings (\cite{RN223}).
Besides, a direct comparison between the ring model and a linear system is missing. Such a comparison would help to delineate the specific advantages offered by lateral connectivity, such as enhanced noise filtering or pattern recognition (\cite{RN221}).

\subsection*{Future work}
Moving forward, several avenues of investigation could further enhance our understanding and application of robustness in visual systems:

\begin{enumerate}
    \item Hierarchical Models: Extending the analysis to hierarchical models of visual processing, including multiple layers that mimic the entire visual pathway, would provide a more comprehensive understanding of robustness across different stages of visual processing (\cite{RN251}).

    \item Comparative Studies: Conducting comparative studies between ring models and network without lateral connections will elucidate the advantages of such connectivity in filtering out abnormal patterns and noise (\cite{RN241}).

    \item Enhanced Biological Models: Incorporating more detailed biological data into spiking network models could improve their realism. This could involve complex neurotransmitter dynamics, dendritic processing, and more accurate replication of neural circuitry (\cite{RN242}).

    \item Artificial Neural Networks Optimization: Applying insights from biological systems to optimize artificial neural networks could lead to more robust machine learning algorithms. Focus should be given to minimizing required sample sizes for training and enhancing network resilience to input perturbations (\cite{RN173}).
\end{enumerate}

In summary, this study lays foundational work for understanding robustness in neural networks through simplified models, providing a pathway towards more robust and adaptive artificial systems that can handle real-world complexities.

\section{Materials \& Methods} \label{sec:materials}
\subsection*{Stead-state ring model}
We use the steady-state rate ring model, which can be viewed as the solution of steady state of a dynamic system.

\begin{equation} \label{eq:steady rate ring model, in materials}
\begin{aligned}
r_{E} &= g(I_{E} + k_{EE}\ast r_{E} - k_{EI}\ast r_{I})\\
r_{I} &= g(I_{I} + k_{IE}\ast r_{E})
\end{aligned}
\end{equation}
$r_{E}$ ($r_{I}$) is a vector denoting the firing rates of excitatory (inhibitory) neuron population. $I_{E}$ ($I_{I}$) represents the excitatory (inhibitory) external input, and $k_{XY}$ ($X,\ Y \in \{E, I\}$) is the connectivity kernel from population $Y$ to $X$, implementing the Gaussian profile.

The activation function $g$ is chosen as the rectified lineaer unit (ReLU), and $\ast$ denotes the circulant convolution operation, enforcing the ring topology, the mathematical formula is as following

\begin{equation} \label{eq:circulant convolution}
x*h = \sum_{k=0}^{N-1}x[k]h[(n-k)\  \text{mod}\ N]
\end{equation}
$x[n]$ and $h[n]$ here are of length $N$.
 
One thing should be careful here is that, $k_{XY}$ is of Gaussian profile, which is symmetric about the axis $\theta=0$. So the picture of $k_{XY}$ is of Gaussian profile based on the angle range from $-90$ degree to $90$ degree. But when we do the convolution, $k_{XY}$ should be adjusted, which start from the angle 0, inversely to the angle $-180$. So the $k$ here need a shift. You see that there is a little bit of symbol abuse going on here and the meaning should be considered in context. 


With small perturbation $\delta I_{E}$ and $\delta I_{I}$, we get the perturbed system

\begin{equation} \label{eq:perturbed rate, in materials}
\begin{aligned}
\delta r_{E} &= g^{\prime}_{E}\odot(\delta I_{E} + k_{EE} * \delta r_{E} - k_{EI} * \delta r_{I})\\
\delta r_{I} &= g^{\prime}_{I}\odot(\delta I_{I} + k_{IE} * \delta r_{E})\\
\end{aligned}
\end{equation}
For $g$ here is ReLU, the element of $g^{\prime}$ value $0$ or $1$ depending on whether the corresponding neuron is active. 
For convenience, we encode the information of $g^{\prime}_{X}$ in a matrix $G_{X}$ and get the matrix-vector form of the perturbed system.

\begin{equation} \label{eq:matrix form, in materials}
\begin{aligned}
\delta r_{E} &= G_{E}(\delta I_{E} + K_{EE}\delta r_{E} - K_{EI} \delta r_{I})\\
\delta r_{I} &= G_{I}(\delta I_{I} + K_{IE}\delta r_{E})\\
\end{aligned}
\end{equation}
$K_{XY}$ is a circulant comes from the lateral connection $k_{XY}$, the first row of it is taken from $k_{XY}$, discretized from angle $0$ to $-180$. Since the kernel $k_{XY}$ is symmetric about $0$, the generated matrix is a symmetric matrix.

\subsection*{Eigenvectors when all neurons are active}
Perform Fourier transformation on both sides of the equation with respect to neuronal positions on the ring

\begin{equation} \label{FT, in materials}
    \begin{aligned}
        \hat{\delta r}_{E} &= \hat{\delta} I_{E} + \hat{k}_{EE} \odot \hat{\delta r}_{E} - \hat{k}_{EI} \odot \hat{\delta r}_{I}\\
        \hat{\delta r}_{I} &= \hat{\delta I}_{I} + \hat{k}_{IE} \odot \hat{\delta r}_{E}\\
    \end{aligned}
\end{equation}
Here the Fourier transformation is in the form
\begin{equation}
    \hat{x} = \sum_{n=<N>} x[n] exp(-ik\omega_{0}n), \omega_{0} =\frac{2\pi}{N}
\end{equation}

The perturbed system can be solved now directly in the frequency space and the solution is
\begin{equation}
\begin{aligned}
    \hat{\delta r}_{E} &= (\hat{\delta I}_{E} - \hat{k}_{EI}\odot\hat{\delta I}_{I})(1- \hat{k}_{EE} + \hat{k}_{EI}\odot\hat{k}_{IE})^{-1} \\
    \hat{\delta r}_{I} &= (\hat{k}_{IE}\odot\hat{\delta I}_{E} + \hat{\delta I}_{I} - \hat{k}_{EE}\odot\hat{\delta I}_{I})(1- \hat{k}_{EE} + \hat{k}_{EI}\odot\hat{k}_{IE})^{-1}
\end{aligned}
\end{equation}
The singular values are given by $(1- \hat{k}_{EE} + \hat{k}_{EI}\odot\hat{k}_{IE})^{-1}$, with the frequency order.
As we have mentioned before, how the singular values against frequency can be classified to several cases. Here we give the specific classification basis when the kernels are give as 

\begin{equation}
k_{XY} = \alpha_{XY}e^{-x^{2}/2\sigma_{XY}^{2}}
\end{equation}
$X, Y\in \{E, I\}$. With these Gaussian kernels, we have them in frequency space as 
\begin{equation}
\hat{k}_{XY}(\xi) = \frac{N}{T}\alpha_{XY}\sqrt{2\pi}\sigma_{XY}e^{-2\pi^{2}\sigma_{XY}^{2}\xi^{2}/T^{2}},
\end{equation}
where $N$ is the number of neurons, $T$ is the period, which takes $180$ here.

Denote $\tilde{\alpha_{XY}} = \frac{N}{T}\alpha_{XY}$, we care about the the change of eigenvalues related to the kernels parameters. We have
\begin{equation}
\begin{aligned}
\hat{h}_{0} &=
1- \hat{k}_{EE} + \hat{k}_{EI}\hat{k}_{IE}  \\
&=
1 -
\tilde{\alpha}_{EE}\sqrt{2\pi}\sigma_{EE}e^{-2\pi^{2}\sigma_{EE}^{2}\xi^{2}/T^{2}} + 
2\pi\tilde{\alpha}_{EI}\tilde{\alpha}_{IE}\sigma_{EI}\sigma_{IE}e^{-2\pi^{2}(\sigma_{EI}^{2}+ \sigma_{IE}^{2})\xi^{2}/T^{2}} \\
\end{aligned}
\end{equation}
There are two key quantities here.

\begin{itemize}
    \item 
    $(\sigma_{EI}^{2} + \sigma_{IE}^{2})/\sigma_{EE}^{2}$. We can view $\sigma_{EI}^{2} + \sigma_{IE}^{2}$ as the scope for the recurrent inhibitory lateral connection and $\sigma_{EE}$ for the excitatory one. So this quantity determine which part has a wider scope.
    \item    $(\tilde{\alpha}_{EE}\sigma_{EE}^{3})/(\tilde{\alpha}_{EI}\tilde{\alpha}_{IE}\sigma_{EI}\sigma_{IE}(\sigma_{IE}^{2} + \sigma_{EI}^{2}))$ determines whether the zero frequency is enhanced/suppressed the most in the most left/right case.
\end{itemize}

\subsection*{Gabor filters}
We are interested in how the model is sensitive to image changes and here is a gap from a signal to an image, or inverse. We bridge the gap by gabor filters.
A gabor filter is constructed as
\begin{equation}
\mathcal{F}_{\mathcal{G}} = 
\begin{bmatrix}
g_1 \\
g_2 \\
\vdots \\
g_n
\end{bmatrix}
\end{equation}
where $g_{i} = g^{(i-1)\omega}, i=1, 2\cdots n$, ($\omega = 2\pi/n$) are row vectors, generated by flattening discretized $2$-d gabor functions with the expression
\begin{equation}
g^{\theta} = g(x,y; A, \lambda, \theta, \psi, \sigma, \gamma) = 
A\exp(-\frac{x'^{2} + \gamma^{2} y'^{2}}{2\sigma^{2}})\cos(2\pi fx'+\psi) 
\end{equation}
where 
\begin{equation}
\begin{aligned}
x' &=  x \cos(\theta) + y \sin(\theta)\\
y' &= -x \sin(\theta) + y \cos(\theta)
\end{aligned}
\end{equation}

\subsection*{I\&F neuronal ring model}

In this work, we consider a fully connected network with conductance-based, integrate-and-fire neuron (\cite{obeid2021stabilized}). The population consists of $N=600$ neurons, each labeled by its orientation preference, $\theta_k = 0.3k$ degrees, which forms a ring. The dynamics of each neuron in the network is modeled by the leaky integrate-and-fire equation:
\begin{equation}
    \tau_{m}\frac{dV}{dt} = -(V - R_{L}) + \frac{g_{E}}{g_{L}}(R_{E} - V) + \frac{g_{I}}{g_{L}}(R_{I} - V) 
\end{equation}
where $\tau_m$ denotes the time constant, $g_L$ is the leak conductance, $g_E$ and $g_I$ are the time dependent excitatory and inhibitory conductances, and $R_L, R_E, R_I$ are reversal potentials. When the neuron's membrane potential reaches threshold $V_{th}$, the neuron generates a spike and the membrane potential returns to the resting potential$V_{rest}$ and remains at the resting potential until the end of refractory period $\tau_{ref}$. Biophysical parameters are used:$g_L=10nS, R_L=-70mV, R_E=0mV, R_I=-80mV, V_{th}=-50mV, V_{rest}=-56mV, \tau_m=15ms, \tau_{ref}=0ms$. For any neuron $n$ of type $X \in \{E,I\}, g_E, g_I \geq 0$ are its excitatory and inhibitory conductances governed by
\begin{equation}
      \frac{dg_{E}}{dt}= -g_{E}/\tau_E + \sum^{N_{E}}_{b=1}\sum_{j}W_{XE}\delta(t - t_{Ej})
  +\sum_k g_{X,ext} \delta(t-t_{ext,k})
\label{eq:g_E}
\end{equation}
\begin{equation}
    \frac{dg_{I}}{dt}= -g_{I}/\tau_{I} + \sum^{N_{I}}_{b=1}\sum_{j}W_{XI}\delta(t - t_{Ij})
\label{eq:g_I}
\end{equation}
where $\tau_E = 3ms$ and $\tau_I = 3ms$ are decay rates for excitatory and inhibitory conductances respectively.And $W_{XY} = w*g_{XY}$, where$w=A+B*exp(-\frac{(\theta_{pre}-\theta_{post})^2}{2\sigma_{ori}^2})$, and $\theta_{pre}$ is reference angle for pre-synaptic neurons, $\theta_{post}$ is reference angle for post-synaptic neurons, $\sigma_{ori}$is the width of Gaussian kernel, $A,B$ are constants satisfying $A+B=1$. Synaptic inputs from other neurons within the network are described in the second terms on the right sides of equations (\ref{eq:g_E}) and (\ref{eq:g_I}) $t_{Ej}, t_{Ij}$are the spike times of all the E- and I-neurons pre-synaptic to neuron n. And external synaptic input is described by a Poisson sequence with rates $r_{ext}$ that arrives at neuron n in $ t_{ext,k}$. And $\delta(\cdot)$ is the dirac delta function indicating an instantaneous jump of conductance $g_E$ or $g_I$ upon the arrival of an E- or I- or external spike, with amplitude equal to $g_E, g_I, g_{ext}$ respectively. In addition, our poissongroups and E-, I- neurons are all one-to-one connected(i.e. we have $2N$ Poisson neurons), and the rates of the Poisson groups of neurons with corresponding preferential angles satisfying the following equation
\begin{equation}
    r_{ext} = A*cos(2\pi\theta f)+C
\end{equation}
Where $A$ represents the amplitude of the fluctuation, $f$ represents the frequency of the fluctuation, $\theta$ is the preference angle of the neuron, and $C$ is the strength of the base frequency inputs. This equation describes how the input changes spatially, i.e. we can generate stimuli of different spatial frequencies. At the same time, we ensure that the mean of the input is constant and the variance of the input can be adjusted, i.e., the product of $r_{ext}$ and $g_{ext}$ should be constant.

If we consider the mean-driven regime mentioned above, we need:
$$ N \rightarrow +\infty, \ \ g_{XY} \rightarrow 0,\ \   N_{input} \rightarrow +\infty $$
which means we tune the network driven primarily by the mean of the inputs rather than fluctuations, i.e., the input variance tends to 0. With such parameters, the network will enjoy even less randomness, which reduces the fluctuations of the network while making it closer to our theoretical analysis.

\subsection*{Measurement of spiking neuron ring model on different frequency inputs}

Our goal is to test changes in the response of the ring model to different frequency inputs under different connectivity conditions (i.e., connection strengths and spatial extents between excitatory and inhibitory neurons). Therefore, we chose to benchmark the response of the ring model to different frequencies without any connections to test the variation due to various connections. Various measurements are accomplished through the following process:
\begin{enumerate}
    \item Vary the fluctuation amplitude $A$ when there is no connection and test the change in the issuance rate.
    \item At the same time, take the discrete Fourier transform of the change in the response and the change in the input current to obtain $\frac{\hat{\Delta r}}{\hat{\Delta I}}$.
    \item Change the fluctuation amplitude $A$ again, but with the corresponding connection and test the change in the issuance rate.
    \item Take a discrete Fourier transform of the new change in issuance rate versus the change in input current to get $\frac{\hat{\Delta r^\prime}}{\hat{\Delta I^\prime}}$.
    \item Comparing the two ratios, that is, we end up with $ \frac{\hat{\Delta r^\prime}}{\hat{\Delta r}}$.
    
\end{enumerate}
Therefore, in our experiments we mainly examine the relationship between $A$ and $\frac{\hat{\Delta r^\prime}}{\hat{\Delta r}}$ as the subject of analysis, which is similar to that analyzed by the steady-state rate model.

\subsection*{Parameter correspondence in mean-driven regime}

Since the steady-state rate model is a dimensionless model, we need to consider its parameteric correspondence to the actual model with the following equations:
\begin{equation}
    r_E = g(I_E+k_{EE}*r_E-k_{EI}*r_I)
\end{equation}
\begin{equation}
    r_I = g(I_I+k_{IE}*r_E)
\end{equation}
where $k_{XY}=\alpha_{XY}exp(-x^2/(2\sigma^2))$. we consider to fix $g=1$ and set $r_E,r_I$ to be same as experimental results with unit Hz, then we need calculate $I_E,I_I,\alpha_{XY}$ in the steady-state rate model.


We will complete the correspondence of the parameters by the following process:
\begin{enumerate}
    \item Determine the correspondence of $I_E,I_I$ in the absence of any connection: $r_E = I_E=kI_E^{T}+b$ where $I_E^T$ is the current measured in the experiment.
    \item We compute $\alpha_{EE}$ when there is only an E to E coupling. We consider $r_E = I_E+k_{EE}*r_E$, and we have$$\alpha_{EE}=\frac{\bar{r_E}-(kI_E^T+b)}{\bar{r_E}\int_0^{180}e^{-x^2/(2\sigma^2)}dx}$$.
    \item We compute $\alpha_{IE}$ when there is only an I to E coupling. We consider $r_E = I_E-k_{IE}*r_I$, and we have$$\alpha_{IE}=\frac{\bar{r_E}-(kI_E^T+b)}{\bar{r_I}\int_0^{180}e^{-x^2/(2\sigma^2)}dx}$$.
    \item We compute $\alpha_{EI}$ when there is only an E to I coupling. We consider $r_I = I_I+k_{EI}*r_E$, and we have$$\alpha_{EI}=\frac{\bar{r_I}-(kI_I^T+b)}{\bar{r_E}\int_0^{180}e^{-x^2/(2\sigma^2)}dx}$$.
\end{enumerate}

Finally, We can reasonably compare the steady-state rate model with the distribution model.

\section{Appendix} \label{sec:appendix}

\subsection*{Gap between steady-state rate models and spiking neuron models}






In the analysis of the steady-state rate model, we can see that the model does not enhance or suppress the response to the high-frequency part. This is because $k_{EE}*r_E, k_{EI}*r_I, k_{IE}*r_E$ are constants at high frequencies, i.e., they do not have high-frequency components. Based on this observation, we can analyze the similar parts of the spiking neuron model $\frac{g_{E}}{g_{L}}(R_{E} - V) , \frac{g_{I}}{g_{L}}(R_{I} - V)$, and analyze whether they change in the high-frequency part.

\begin{figure}[h!]
    \centering
    \includegraphics[width=0.6\textwidth]{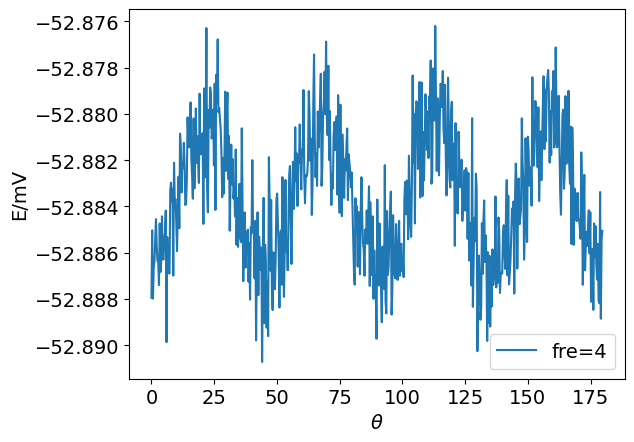}
    \caption{Fluctuation of potentials. This figure shows the average value of a neuron's potential over time, which implies that the potential of a neuron fluctuates as a result of inputs.}
    \label{fig:v flu}
\end{figure}

Through experiments, we found that, unlike the steady-state rate model, the potential fluctuates in the spiking neuron model by fluctuating inputs as shown in figure\ref{fig:v flu}. We believe that this fluctuation in potential affects the model's performance in the high-frequency part. That is, models with lateral connections will somewhat enhance or weaken the high-frequency part, compared to models without connections.

Based on the above understanding, we will consider a more simplified model to analyze how fluctuations in potential can have an effect in a model with lateral connections. We consider the equation in the steady state, i.e., we regard $g_E, g_I, g_ext, g_in$ (with spatial fluctuations) as constants for analysis, and we can get the following equation:
\begin{align}
\frac{dv}{dt} &= -a(v-\frac{b}{a})
\end{align}

This is a simplification of the above equation. At the same time, we consider the average potential during the two firing processes, and even more simply, we consider the average potential from $t=0, v=v_0$ to the first time the firing potential is reached $t=t_1, v=v_1$ (this moment is related to a and b), we get the following result:
\begin{align}
\bar{v}&=\frac{b}{a}+\frac{v_0-v_1}{ln\frac{av_0-b}{av_1-b}}\\
t_1 &= \frac{1}{a}ln\frac{av_0-b}{av_1-b}
\end{align}

That is to say, $\bar{v}$ will vary with the connection parameters and the firing rate, where $t_1$ is affected by the firing rate. On this basis, we give the values of a and b :

\begin{equation}
a=\frac{g_L+g_{ext}+g_E+g_I}{\tau_mg_L},
b=\frac{R_Lg_L+R_E(g_{ext}+g_E)+R_Ig_I}{\tau_mg_L}
\end{equation}

\begin{figure}[h!]
    \begin{minipage}{0.3\textwidth}
    \centering
    \includegraphics[width=\textwidth]{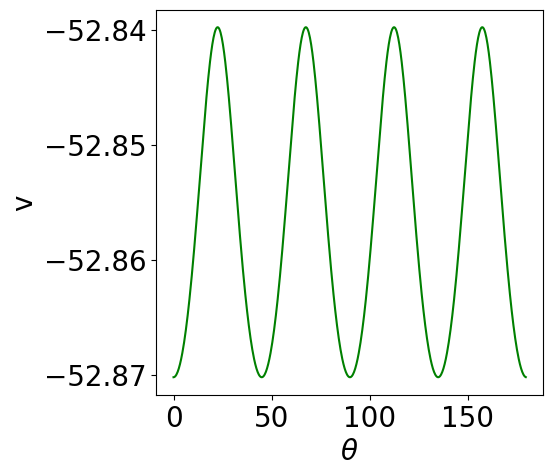}
    \end{minipage}
    \hfill
    \begin{minipage}{0.3\textwidth}
    \centering
    \includegraphics[width=\textwidth]{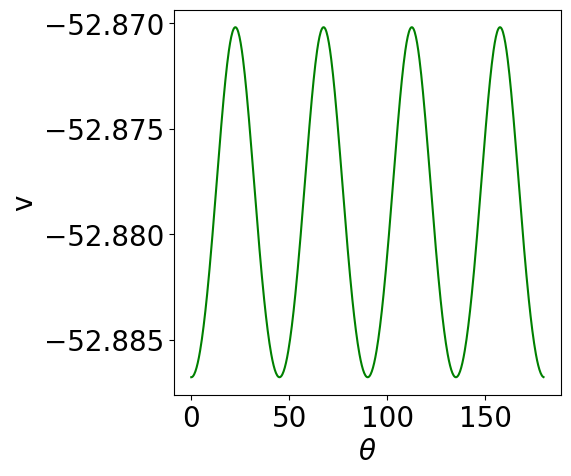}
    \end{minipage}
    \hfill
    \begin{minipage}{0.3\textwidth}
    \centering
    \includegraphics[width=\textwidth]{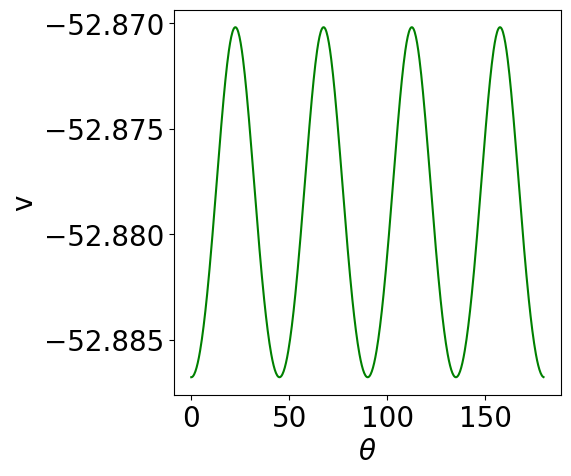}
    \end{minipage}
    \caption{Theory result of $\bar{v}$: (left)Benchmark: C=12, amp=1.5, $g_E=g_I=0$, (middle)Increase E connection: $g_E=3$, (right)Increase I connection: $g_I=3$. }
    \label{fig:flu theory}
\end{figure}

\begin{figure}[h!]
    \centering
    \includegraphics[width=0.6\textwidth]{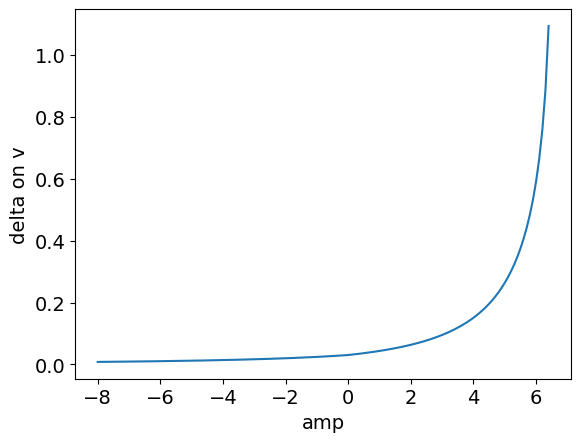}
    \caption{Effect of different connections on the magnitude of fluctuations. And amp refers to the amplitude of $g_E$ and $g_I$, where a negative number indicates the strength of $g_E$ and a positive number indicates the strength of $g_I$. This implies that inhibitory connections lead to stronger fluctuations than excitatory connections.}
    \label{fig:flu all}
\end{figure}

Then, we consider the input as:
\begin{align}
g_{ext} = C + amp*cos(2\pi\theta*fre)
\end{align}
Under the condition of fixed input strength and fluctuation strength, we found that there is inconsistency on potential spatially when increasing the values of $g_E, g_I$ as shown in figure\ref{fig:flu theory}.
We can find that the impact of the fluctuation amplitude is inconsistent, and more specifically, we analyzed this inconsistency as shown in figure\ref{fig:flu all}.
The negative part represents the increase in $g_E$, and the positive part represents the increase in $g_I$. Based on the observations, we can find that the model's impact on $g_E, g_I$ is inconsistent, that is, it should be an nonlinear change rather than a linear change. That means, at the same strength, the I connection naturally brings more fluctuation changes than the E connection in the high-frequency part.

\subsection*{Fluctuation-driven regime} \label{sec:fluctuation driven}

We first present the key parameters adopted under the mean-driven regime. Unless specifically mentioned, they are consistent with the parameters selected earlier in the text. Here, we only repeat some of the key parameters: we choose the number of neurons $N=240$, set $N_{\text{input}}=1$, and the refractory period $\tau_{\text{ref}}=3ms$. The settings for the fluctuation-driven regime will introduce more noise into our model and bring about greater nonlinearity. We will first demonstrate the results under the fluctuation-driven regime and analyze the differences between mean-driven regime and fluctuation-driven regime.

\subsubsection*{Results in the Fluctuation-Driven Region}

\begin{figure}[h!]

    \begin{minipage}{0.3\textwidth}
    \centering
    \includegraphics[width=\textwidth]{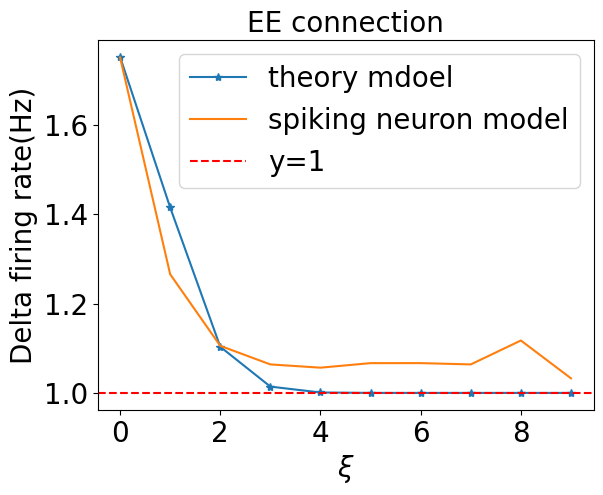}
    \end{minipage}
    \hfill
    \begin{minipage}{0.3\textwidth}
    \centering
    \includegraphics[width=\textwidth]{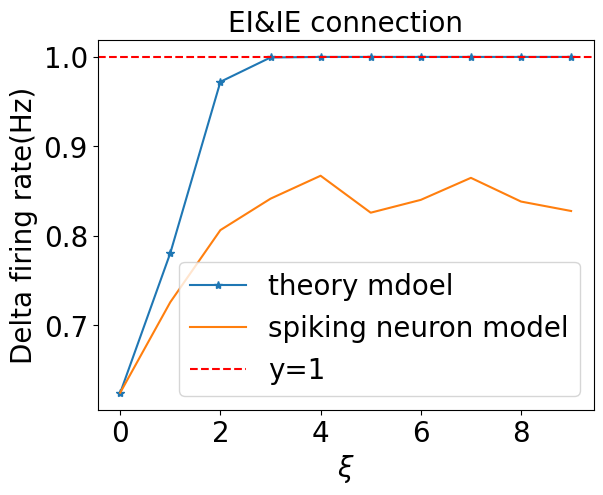}
    \end{minipage}
    \hfill
    \begin{minipage}{0.3\textwidth}
    \centering
    \includegraphics[width=\textwidth]{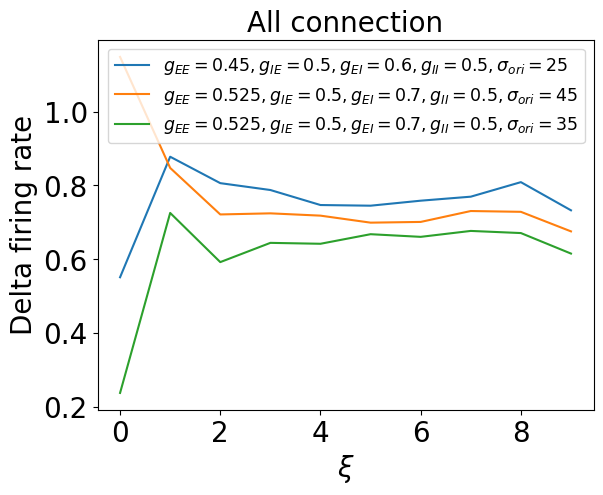}
    \end{minipage}
    \caption{Gain curves of spiking neuron model(fluctuation-driven regime). The figure on the left shows the results with only recurrent excitation connections. The middle figure shows the results with only recurrent inhibition. The right figure demonstrates the results with full connectivity. Due to the complexity of full connectivity, its quantitative comparison with theoretical models is difficult. Therefore, we show the results of multiple qualitative full-connectivity cases without comparing them to the theoretical model.}
    \label{fig:flutuation}
\end{figure}

In fluctuation-driven regime, we obtain results that are qualitatively consistent with the theoretical results. 
As shown in the Figure \ref{fig:flutuation}, when no connections were present, the ring model responded uniformly to all frequencies. With only recurrently excitatory connections, the ring model enhanced the low-frequency response. In contrast, when only recurrently inhibitory connections were present, the ring model suppressed the low-frequency response. And when all connections are present, the model demonstrates a choice of specific frequency.

\begin{figure}[!htp]
  \centering
  \includegraphics[width=1\textwidth]{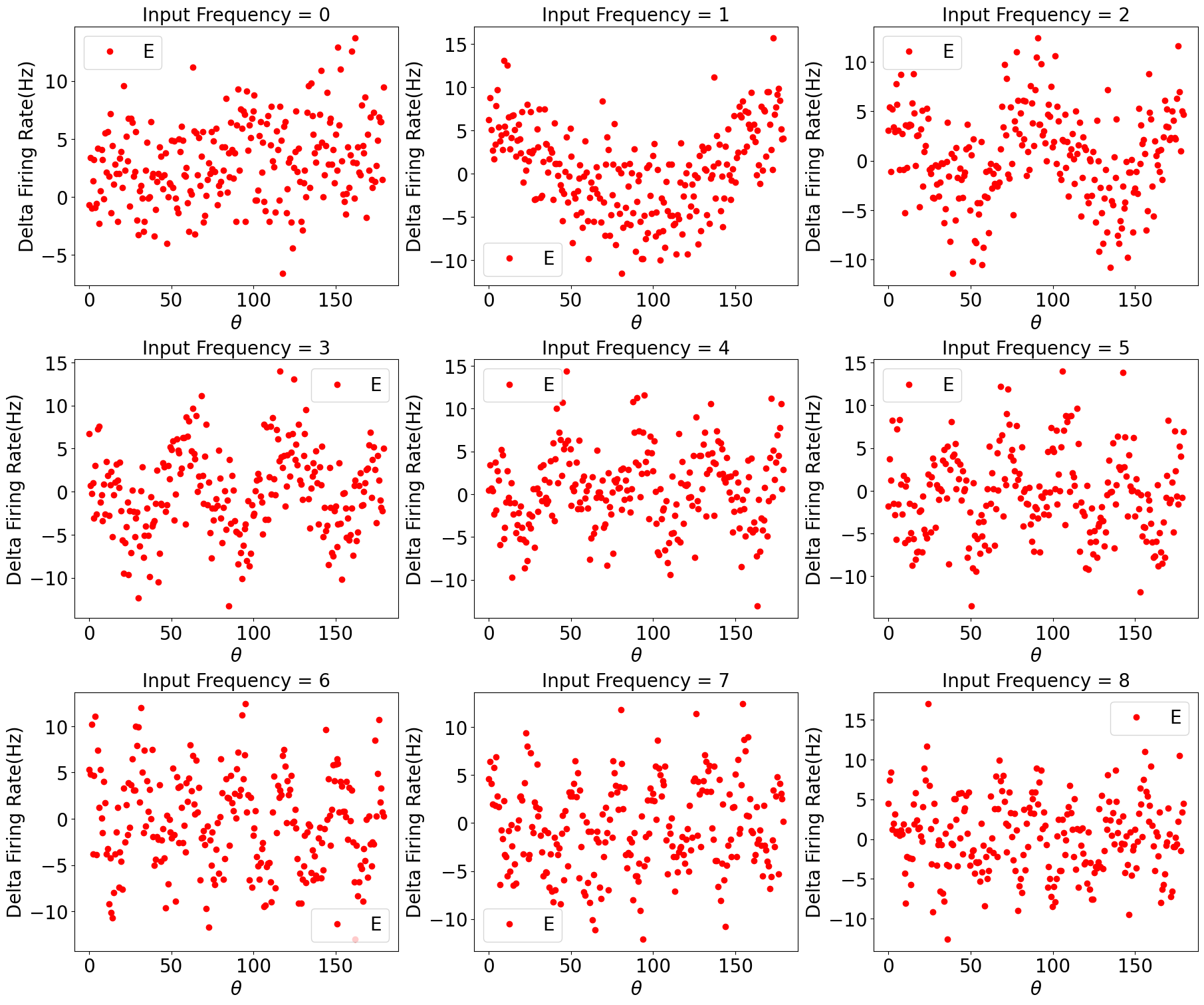}  
  \caption{The effect of perturbation on firing rates. This figure shows the change in frequency space brought about by different frequency inputs in the case of full connectivity}
  \label{fig:rates flu}
\end{figure}

Although these results are qualitatively consistent, quantitatively we can still find significant differences between fluctuation-driven regime and mean-driven regime:
\begin{enumerate}
    \item \textbf{Difference in fluctuation intensity}: From the response curves, it is evident that the mean-driven regime has almost no fluctuations, whereas the fluctuation-driven rigime exhibits strong fluctuations, as shown in Figure \ref{fig:rates flu}.
    \item \textbf{Difference in high-frequency responses}: The high-frequency responses in the mean-driven regime are primarily caused by potential fluctuations, while in the fluctuation-driven regime, they arise from more complex factors, which we will analyze in detail later.
    \item \textbf{Difficulty in adjusting full connectivity}: In the mean-driven regime, selecting specific frequencies is relatively easy, whereas in the fluctuation-driven regime, more parameters need to be changed, such as the width of connections between neurons $\sigma_{\text{ori}}$.
\end{enumerate}
Due to the complexity of the spiking neuron model, more factors need to be considered, such as the differences between mean and fluctuation-driven regimes, nonlinearity in the gain curves, and changes in the Gaussian kernel. These details still differ from the steady-state rate model, but we will discuss them further in the following text. However, these differences do not affect the key characteristics, and from a qualitative perspective, the responses are still consistent with the steady-state rate model.

\subsubsection*{Impact of the Neuronal Population Gain Curve}

\begin{figure}[!htp]
  \centering
  \includegraphics[width=1\textwidth]{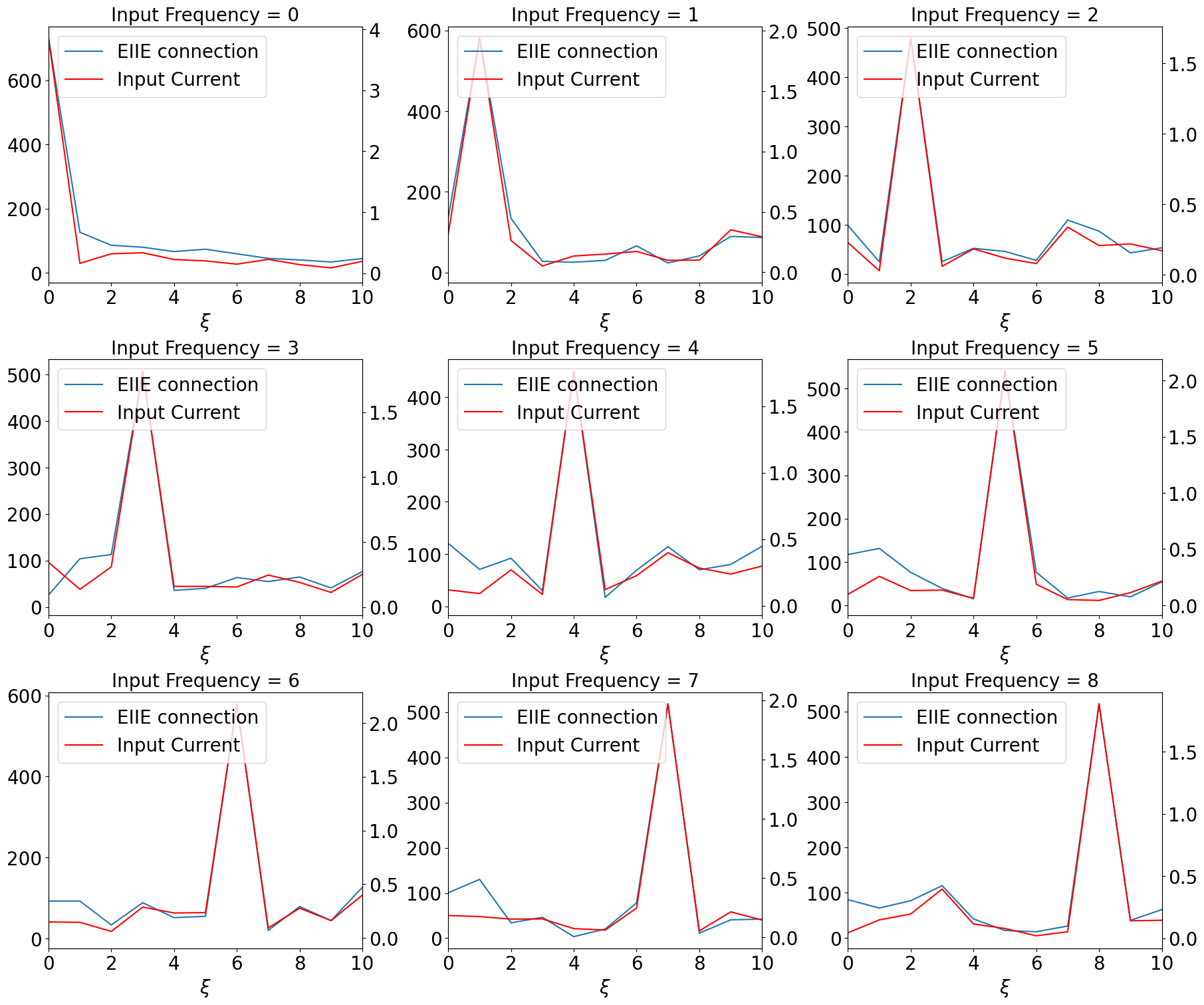}  
  \caption{The effect of perturbation on firing rates in frequency space. This figure shows the change in frequency space brought about by different frequency inputs in the case of full connectivity}
  \label{fig:fre flu}
\end{figure}

\begin{figure}[!htp]
  \centering
  \includegraphics[width=0.75\textwidth]{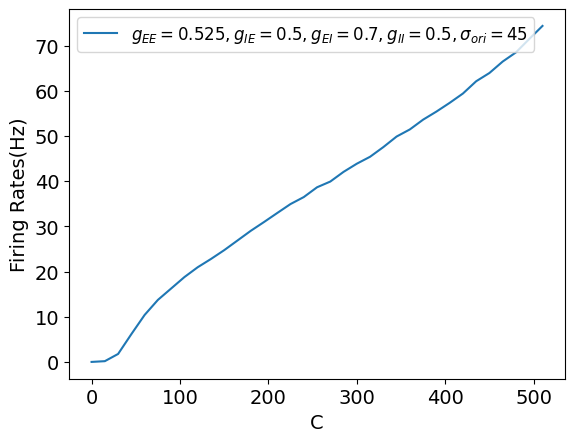}  
  \caption{Gain curve of the fully connected ring model population.}
  \label{fig:gain flu}
\end{figure}

Further, we observe the firing rates in the frequency space, as shown in Figure \ref{fig:fre flu}. Compared to the mean-driven regime, the ring model shows a notably enhanced response at the fundamental frequency.
In response to the fundamental frequency, analyzing the neuronal population gain curve can help us understand this phenomenon. We have plotted the gain curve of the neuronal population, as shown in Figure \ref{fig:gain flu}. We find that in the fluctuation-driven regime, it is more challenging to find an approximately linear region in the neuron's gain curve, unlike in the mean-driven regime.
When the gain curve is linear, there is no significant enhancement of the fundamental frequency; however, in the nonlinear region, the response strength at the fundamental frequency increases.

\subsubsection*{Responses to Inputs with Different Variances}

\begin{figure}[htbp]
    \centering
    \includegraphics[width=0.75\textwidth]{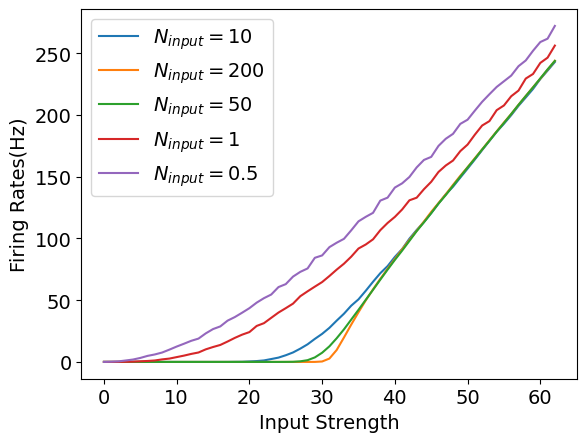}
    \caption{The effect of the variance of the inputs. A smaller $N_{input}$ corresponds to a larger variance.}
    \label{fig:gain flu single}
\end{figure}

Another difference between fluctuation-driven and mean-driven regimes arises from the impact of input variance on the model. When analyzing this impact, we need to consider the gain curve of individual neurons, as depicted in Figure \ref{fig:gain flu single}. Figure \ref{fig:gain flu single} shows the effects of inputs with different variances on the model, and in the fluctuation-driven regime, we can consider the network's input to a particular neuron as an external input, which aligns with Figure \ref{fig:gain flu single} to some extent. Additionally, inhibitory and excitatory inputs within the network bring about this variance change, while in the mean-driven regime, since both the input and network variances are relatively small, this impact can be negligible.
However, the internal inputs within the network are not standard Poisson processes, differing from the external inputs. Therefore, when considering the impact of internal inputs, analysis of higher-order statistics might be necessary. Despite this, the differences exhibited at the variance level can help us understand the distinctions between the fluctuation-driven and mean-driven regimes to some extent.

Through these analyses, we delve deeper into the complex dynamics of the neuronal network under different driving regime, highlighting the complex balance between network structure, connectivity, and input characteristics. The nuanced differences between fluctuation-driven and mean-driven regimes underscore the sensitivity of neuronal responses to changes in network parameters and external conditions, offering insights into the adaptability and functionality of neural circuits in varying operational states.

\section{Acknowledgement} 
This work is supported from the National Key R\&D Program of China Grant No. 2022YFA1008200 (Y. Z.), the National Natural Science Foundation of China Grant No. 12101402 (Y. Z.), the Lingang Laboratory Grant No.LG-QS-202202-08 (Y. Z.), Shanghai Municipal of Science and Technology Major Project No. 2021SHZDZX0102 (Y. Z.), the Science and Technology Innovation 2030 —— Brain Science and Brain-Inspired Intelligence Project No. 2021ZD0201301 (W. D.), the National Natural Science Foundation of China No. 12201125 (W. D.)  and  the Science and Technology Committee of Shanghai Municipality No. 22YF1403300 (W. D.).

\printbibliography

\end{document}